\documentclass[fleqn, usenatbib, twocolumns]{mnras}

\usepackage{newtxtext,newtxmath}

\usepackage[T1]{fontenc}
\usepackage{ae,aecompl}

\usepackage{graphicx}	
\usepackage{amsmath}	
\usepackage{amssymb}	
\usepackage{subfig}
\usepackage{gensymb}
\usepackage{txfonts}
\usepackage{tablefootnote}
\usepackage{multirow}
\usepackage{longtable}
\usepackage{threeparttable}

\title[Restarted morphology in hard X-ray selected GRG]{Hard X-ray selected giant radio galaxies -- II. Morphological evidence of restarted radio activity}

\author[G. Bruni et al.]{G. Bruni,$^{1}$\thanks{Contact e-mail: \href{mailto:gabriele.bruni@inaf.it}{gabriele.bruni@inaf.it}}
    F. Panessa,$^{1}$
    L. Bassani,$^{2}$
    D. Dallacasa,$^{3,4}$
    T. Venturi,$^{4}$
    L. Saripalli,$^{5}$
    \newauthor
    M. Brienza,$^{3,4}$
    L. Hern\'andez-Garc\'ia,$^{6}$
    E. Chiaraluce,$^{1}$
    F. Ursini,$^{2}$
    A. Bazzano,$^{1}$
    \newauthor
    A. Malizia,$^{2}$
    and
    P. Ubertini$^{1}$
\\
    $^{1}$INAF - Istituto di Astrofisica e Planetologia Spaziali, via Fosso del Cavaliere 100, I-00133 Roma, Italy \\
    $^{2}$INAF - Osservatorio di astrofisica e scienza dello spazio di Bologna, Via Piero Gobetti 93/3, 40129 Bologna, Italy \\
    $^{3}$DiFA - Dipartimento di Fisica e Astronomia, Universit\`a di Bologna, via P. Gobetti 93/2, I-40129 Bologna, Italy \\
    $^{4}$INAF - Istituto di Radioastronomia, via Piero Gobetti 101, I-40129 Bologna, Italy \\
    $^{5}$Raman Research Institute, C. V. Raman Avenue, Sadashivanagar, Bangalore 560080, India \\
    $^{6}$IFA - Instituto de F\'isica y Astronom\'ia, Facultad de Ciencias, Universidad de Valpara\'iso, Gran Breta\~na 1111, Playa Ancha, Valpara\'iso, Chile \\}

\date{Accepted XXX. Received YYY; in original form ZZZ}

\pubyear{2015}

\begin{document}
\label{firstpage}
\pagerange{\pageref{firstpage}--\pageref{lastpage}}
\maketitle


\begin{abstract}
About 6\% of Radio Galaxies (RG) can reach linear sizes larger than 0.7 Mpc, and are then classified as Giant Radio Galaxies (GRG). The conditions that make possible the formation of such big structures is still not clear - either core accretion properties or environmental factors. Recent studies have shown that GRG can be up to four times more abundant in hard X-ray selected (i.e. from \emph{INTEGRAL}/IBIS and \emph{Swift}/BAT at $>$20 keV) RG samples. Moreover, a high fraction of young radio sources found in their cores suggests a recently restarted activity, as suggested from the discrepancy between the measured jet and lobes power, with respect to the one expected from core X-ray luminosity. 
Here we present a radio morphological study of a sample of 15 hard X-ray selected GRG, discussing low-frequency images from our GMRT campaign complemented with others from the literature: among them, 7/15 show evidence of restarted radio activity either in the form of double-double/X-shaped morphology, or as a cocoon emission embedding more recent jets. This, together with the objects from this sample already found hosting a young radio source in their core, suggests that at least 13 over 15 of these hard X-ray selected GRG show features which are consistent with the possibility of restarted radio activity. 
\end{abstract}

\begin{keywords}
galaxies: active -- galaxies: nuclei -- galaxies: jets -- galaxies: Seyfert -- X-rays: galaxies -- radio continuum: galaxies.
\end{keywords}


\section{Introduction}

Active Galactic Nuclei (AGN) are among the most extreme astrophysical laboratories of the Universe, where the presence of a strong gravitational field from the central supermassive black hole (SMBH), and ionized matter falling onto that, can give rise to strong magnetic fields and relativistic outflows called jets. The synchrotron radiation emitted from the latter is easily detectable in the radio band (from $\sim$50 MHz to $\sim$100 GHz), and define the sub-class called "radio-loud" AGN, constituting $\sim$10\% of the total AGN population. During the past decades, there has been growing evidence that the radio activity - i.e. jet production - of an AGN is of episodic nature. Examples of newborn radio jets, embedded in remnant emission from past activity have been found (see \citealt{2009BASI...37...63S} and references therein), as well as sources with two sets of lobes (named double-double) suggesting two distinct episodes of activities \citep{2000A&AS..146..293S}. In some cases, the newborn jets can have a different axis with respect to the previous ones (e.g. \citealt{2013MNRAS.436..690S}) possibly producing an X-shaped morphology. The estimated life-cycle for sources showing more than one radio phase accounts for an active phase in the range $10^7-10^8$ yr, and a quiescent phase in-between of $10^5-10^7$ yr (see \citealt{2017NatAs...1..596M} and references therein). 
 
Giant Radio Galaxies (GRG, \citealt{1974Natur.250..625W,1999MNRAS.309..100I}), with a linear extent of more than 0.7 Mpc and ages of tens of Myr, are then the ideal laboratories to test the presence of multiple radio-activity phases, and probe their duty cycles. Indeed, some authors suggest that GRG could reach their extreme extent thanks to multiple activity cycle with relatively short quiescent periods \citep{1996MNRAS.279..257S,2005AJ....130..896S}. The number of known GRG has noticeably increased in the last years, thanks to the release of recent GRG catalogues collecting studies of single sources from the literature \citep{2018ApJS..238....9K}, as well as the ongoing new radio surveys at low frequencies ($<$1 GHz) made with the latter generation radio telescopes, as the LOFAR Two-metre Sky Survey (LoTSS, \citealt{2017A&A...598A.104S}). Indeed, their steep-spectrum radio emission makes GRG shine more evidently at low frequencies. Nevertheless, GRG known to date number only about 350, among a total population of thousands of radio galaxies ($\sim$6\%, \citealt{1999MNRAS.309..100I}). 

The role of the environment in the growth of radio galaxies up to the extreme sizes of GRG is still under debate. It has been proposed that GRG could lie in regions of cosmic void, and that can, therefore, more easily expand in the intergalactic medium \citep{2004evn..conf..117M}. Separately, in a radio and optical spectroscopic study of a sample of 19 GRG, \cite{2015MNRAS.449..955M} found support for environmental role in growth to giant sizes, finding GRG axes to be preferentially oriented along regions of lower galaxy density, suggesting that jets that are oriented in directions that are relatively devoid of neighbouring galaxies can lead to giant extents. Eventually, the correspondence of GRG positions with cosmic voids seemed to be disfavoured by works with a larger statistics, making use of a sample of hundreds of objects \citep{2018ApJS..238....9K}.  Recently, \cite{2019arXiv190400409D} compiled a sample of 240 GRG from the LoTSS survey (150 MHz), finding that at redshifts below $z\sim0.4$ at least 18\% of GRGs can be found at the center of galaxy groups or clusters. 

In this work, we discuss the radio morphology of a sample of 15 GRG extracted from a catalogue of radio galaxies selected from the hard X-ray band \citep{2016MNRAS.461.3165B}. 
We adopt the latest cosmological parameters from the \emph{Planck} mission (\citealt{2018arXiv180706209P}), i.e. assuming the base-$\Lambda\rm{CDM}$ cosmology: $H_{0}$=$67.4$ km/s/Mpc, $\Omega_{\rm{m}}$=$0.315$, and $\Omega_{\Lambda}$=$0.685$.


\section{The hard X-ray selected GRG sample}

We are carrying out a multi-wavelength study of a sample of hard X-ray ($>$20 keV) selected GRG, extracted from high energy catalogues (the GRACE\footnote{\href{https://sites.google.com/inaf.it/grace/}{https://sites.google.com/inaf.it/grace/}} project - \emph{Giant RAdio galaxies and their duty CyclE}).
The 15 GRG subject of this work have been extracted from a parent sample of 64 radio galaxies selected via cross-correlation of hard X-ray catalogues and radio surveys. Indeed, \cite{2016MNRAS.461.3165B} produced a catalogue of radio galaxies detected by the \emph{INTEGRAL}/IBIS (Imager on Board the INTEGRAL Satellite) and/or \emph{Swift}/BAT (Burst Alert Telescope), at energies $>$20 keV, identified using the two complementary radio surveys covering the whole sky: the NRAO Very Large Array Sky Survey (NVSS, 1.4 GHz, \citealt{1998AJ....115.1693C}), and the Sydney University Molonglo Sky Survey (SUMSS, 843 MHz, \citealt{1999AJ....117.1578B}). The same authors found that $\sim$25\% of the objects in that sample have a projected linear size $>$0.7 Mpc, and can then be considered GRG (see Tab. \ref{sample}). This fraction is almost four times larger than what previously found in radio galaxies samples selected from the radio ($\sim$6\% in the 3CR catalog, see \citealt{1999MNRAS.309..100I} and 8\% in the 3CRR sample \citep{1983MNRAS.204..151L}. \cite{2018MNRAS.481.4250U} (hereinafter referred as paper I, preceding this work) presented a X-ray follow-up of the sample, finding that in most sources the jet power and the radio luminosity of the lobes are lower than what is expected from the nuclear X-ray emission: this could be the case if the core was undergoing a restarted activity. \cite{2019ApJ...875...88B} built the radio spectra of the cores of these GRG, showing that more than half show a spectrum peaking in the GHz domain, consistent with being young radio sources with an age of the order of tens of kyr, in contrast with the Myr timescale needed to produce the extended lobes: this came out in favor of the restarted scenario proposed by \cite{2018MNRAS.481.4250U}. In this paper we discuss cases of restarted radio morphology, by collecting low frequency radio images either from our dedicated observational campaign or from the literature/surveys.    


\begin{table}
\centering
\caption{Details about data collected for this work. From lower to higher frequencies: TGSS (\citealt{2017A&A...598A..78I}), GMRT (this work), SUMSS (\citealt{1999AJ....117.1578B}), NVSS (\citealt{1998AJ....115.1693C}), ATCA (\citealt{2008arXiv0806.3518S,2013MNRAS.436..690S}). The NVSS and SUMSS surveys together, used to define this sample of GRG, cover the entire sky. }
 \begin{tabular}{ccccccc}
  \hline
    Telescope   &   Frequency   &   Beam axes                   &   Typical RMS  \\
                &   (MHz)       &   (arcsec)                    &   (mJy/beam)   \\
  \hline
    TGSS        &   150         &   25$\times$25                &    5           \\ 
    GMRT        &   325         &   15$\times$7; 27$\times$14   &    0.3         \\
    SUMSS       &   843         &   43$\times$43$\times{csc}\,|\delta|$                &    20          \\
    NVSS        &   1400        &   45$\times$45                &    0.45        \\    
    ATCA        &   1400        &   7.5$\times$4.9              &    0.1         \\
    ATCA        &   2500        &   18$\times$14                &    0.5         \\
  \hline
 \end{tabular}
 \label{observations}
\end{table}



\begin{table*}
\begin{threeparttable}
\caption{The sample of 15 GRG studied in this work. Coordinates of the core from X-ray observations are given (J2000), collected from \protect\cite{2013ApJS..207...19B,2012MNRAS.426.1750M,2016MNRAS.460...19M,2014A&A...565A...2M,2015MNRAS.451.2370M,2012A&A...545A.101P,2014A&A...561A..67P}. In columns 6-9 we mark the evidence of restarted activity for each source. Literature references for restarted activity from previous works is given in column 10: 1) \protect\cite{1998A&A...336..455S}; 2) \protect\cite{2013MNRAS.436..690S}; 3) \protect\cite{2016AcA....66...85W}; 4) \protect\cite{2015MNRAS.451.2370M}; 5) \protect\cite{1992MNRAS.259P..13P}; 6) \protect\cite{2017A&A...603A.131H}; 7) \protect\cite{2019ApJ...875...88B}; 8) \protect\cite{1999ApJ...522..101G}; 9) \protect\cite{2007A&A...474..409G}; 10) \protect\cite{2008arXiv0806.3518S}; 11) \protect\cite{1996MNRAS.279..257S}.}
 \begin{tabular}{ccccccccccccc}
  \hline
    Source          &	ID                  &	RA			            & DEC				        & z			& Cocoon  		& GPS/HFP core      & Double-Double     & X-shaped      &	Reference \\
    (1)			    &	(2)		            & (3)		                & (4)	   	                & (5)		& (6)		    &	(7)	            &   (8)             & (9)           &   (10)       \\
  \hline                        
    J0313+4120	    &   B3 0309+411B	    &	03$^{h}$13$^{m}$01.9$^{s}$	&   +41$^\circ$20'01''		& 0.13	 	& -              & \checkmark       &	-			    &  -            &	7	    \\
    J0318+6829	    &   J0318+684           &	03$^{h}$18$^{m}$19.1$^{s}$	&   +68$^\circ$29'32''	    & 0.09	 	& -              & \checkmark       &	-	            &  \checkmark   &	1,7		\\
	J0709$-$3601    &   PKS 0707--35 	    &	07$^{h}$09$^{m}$13.8$^{s}$	& $-$36$^\circ$01'19''		& 0.11	 	& -              & -                &	-	            &  \checkmark   &	2,11	\\
	J0949+7314      &   4C 73.08		    &	09$^{h}$49$^{m}$46.0$^{s}$	&   +73$^\circ$14'23''		& 0.06	 	& \checkmark     & -                &	-			    &  -            &	3	    \\
	J1147+3501      &   B2 1144+35B		    &	11$^{h}$47$^{m}$22.3$^{s}$	&   +35$^\circ$01'09''		& 0.06		& -              & -                &  \checkmark	    &  -            &	8,9	 	\\
	J1436$-$1613    &   HE 1434--1600   	&	14$^{h}$36$^{m}$49.6$^{s}$	& $-$16$^\circ$13'40''		& 0.14		& -              & -                &	-       	    &  -            &	-		\\
	J1448$-$4008    &   IGR J14488--4008    &	14$^{h}$48$^{m}$50.9$^{s}$	& $-$40$^\circ$08'47''		& 0.12		& -              & \checkmark       &	-			    &  -            &	4,7	    \\
	J1523+6339      &   4C +63.22   	    &	15$^{h}$23$^{m}$45.8$^{s}$	&   +63$^\circ$39'24''		& 0.20		& -              & -                &   -	            &  \checkmark   &	-	    \\
	J1628+5146      &   Mrk 1498  		    &	16$^{h}$28$^{m}$03.8$^{s}$	&   +51$^\circ$46'30''		& 0.05		& -              & \checkmark       &	-			    &  -            &	7	    \\
	J1748$-$2335    &   IGR J17488--2338	&	17$^{h}$48$^{m}$38.9$^{s}$	& $-$23$^\circ$35'26''		& 0.24		& -              & \checkmark       &	-			    &  -            &	7	    \\
	J1723+3418      &   4C +34.47      	    &	17$^{h}$23$^{m}$20.6$^{s}$	&   +34$^\circ$18'00''		& 0.21		& -              & \checkmark       &	-			    &  -            &	7	    \\
	J2018$-$5539    &   PKS 2014--55 	    &	20$^{h}$18$^{m}$01.0$^{s}$	& $-$55$^\circ$39'28''		& 0.06		& -              & -                &  \checkmark	    &  \checkmark   &	10	    \\
	J2042+7508      &   4C +74.26   	    &	20$^{h}$42$^{m}$36.7$^{s}$	&   +75$^\circ$08'02''		& 0.10		& -              & \checkmark       &   -			    &  -            &	5,7	    \\
	J2333$-$2343    &   PKS 2331--240       &	23$^{h}$33$^{m}$55.2$^{s}$	& $-$23$^\circ$43'41''		& 0.05		& -              & -                &   -    	        &  \checkmark \tnote{$\dagger$}   &	6	    \\
	J2359$-$6054    &   PKS 2356--61  	    &	23$^{h}$59$^{m}$03.9$^{s}$	& $-$60$^\circ$54'59''		& 0.09		& -              & \checkmark       &  \checkmark	    &  -            &	7,11	\\
 \hline
 \end{tabular}
 \begin{tablenotes}
    \item[$\dagger$] this source shows a reorientation of almost 90 degrees for the inner jet, with the latter pointing towards the observer's line of sight. Thus, we mark it as X-shaped since it would appear as such when seen from a different angle.
 \end{tablenotes}
 \label{sample}
\end{threeparttable}
\end{table*}


\section{Collected data}

In order to study the radio morphology of these 15 GRG, we collected the best images available at low frequencies either from our GMRT campaign, covering 4/15 sources, or from the literature and surveys, i.e. the TIFR GMRT Sky Survey at 150 MHz (TGSS, \citealt{2017A&A...598A..78I}) and the previously mentionted NVSS and SUMSS (for the remaining 11/15 sources). Details about the obtained images are given in the following (see Tab. \ref{observations} for a summary), while a morphological and spectral analysis is presented in Sec. \ref{radio}.


\subsection{GMRT observations and data reduction}

In May 2014, we carried out GMRT observations for four sources of the sample: B3\,0309+411B,  LCF\,2001\,J0318+684, IGR\,J14488$-$4008, and IGR\,J17488$-$2338 (project code 26\_032). They were taken in snapshot mode at 325 MHz for all sources (total bandwidth 32 MHz), and additionally at 610 MHz for IGR\,J14488$-$4008 only - the latter was already presented in a previous publication \citep{2015MNRAS.451.2370M} so we do not further discuss them here. The on-source observing time was $\sim$4 hours, aiming at an RMS of $\sim$0.2 mJy/beam. Amplitude scale calibration was performed with observations on the well known sources 3C\,286, 3C\,48, and 3C\,147, and suitable phase calibrators near target sources were observed at regular intervals. Data were correlated with the GMRT Software correlator Backend (GSB). Data were reduced with the {\tt{FLAGCAL}} pipeline \citep{2012ExA....33..157P} and once split UV datasets were obtained for each target, imaging and self-calibration were performed in {\tt{AIPS}}\footnote{\href{http://www.aips.nrao.edu/index.shtml}{http://www.aips.nrao.edu/index.shtml}}. Satisfactory images at full resolution were obtained for all sources except for B3\,0309+411B, where the RMS is about ten times higher than expected ($\sim$2 mJy/beam), preventing a detailed study of the diffuse, low surface brightness emission. We could reach an angular resolution of about 15"$\times$8" for J0318+684 and IGR\,J17488-2338, while 27"$\times$14" for IGR\,J14488-4008 (due to the low declination), and a typical RMS of $\sim$0.3 mJy/beam.

One additional source (PKS\,2331$-240$) was observed in a further run in March 2018 (project code 33\_003). It was observed at 150 MHz for 6 hours on-source, with a total bandwidth of 32 MHz, applying  the same strategy as for the previous run. The choice of the observing frequency, different from the 2014 run, was motivated by a possible extended structure visible in the TGSS map, that we aimed to study with a better sensitivity. Due to severe Radio Frequency Interferences (RFI), a bandwidth of only 16 MHz over 32 MHz could be used, so the RMS in the final image was limited to $\sim$1.5 mJy/beam - which is anyway more than a factor of 2 improvement with respect to the TGSS ($\sim$3.5 mJy/beam). The full resolution image has a beam of 25.4"$\times$15.2". For the data reduction of this source, performed more recently with respect to the one for the previous run, we could make use of the {\tt SPAM} pipeline (\citealt{2014ascl.soft08006I,2017A&A...598A..78I}, the same used for the TGSS survey data processing).


\subsection{Ancillary data from surveys and the literature}

In 2017, the TGSS radio survey was published (TGSS ADR1, \citealt{2017A&A...598A..78I}; sensitivity: <5 mJy/beam; angular resolution: 25"). Images from TGSS could be collected for our entire sample, except for PKS 2356--61 which is below the declination limit of the survey, and for which we present the SUMSS image. For two sources (B2 1144+35B and HE 1434--1600) we report the NVSS image, since the extended emission is better detected than in the TGSS one. 

In addition to the survey data, we could also collect images from previous studies for three sources: PKS 0707--35 and PKS 2014--55, for which the ATCA image at 2.5 GHz and 1.4 GHz from \cite{2013MNRAS.436..690S} and \cite{2008arXiv0806.3518S} (respectively) is reported, and 4C\,73.08, for which the combined NVSS+Effelsberg image at 1.4 GHz was kindly provided by M. Wezgowiec (previously published in \citealt{2016AcA....66...85W}). 


\subsection{Position of the cores from X-rays observations}

We collected the X-ray position of the cores for the 15 GRG, available from previous studies in the literature (typical error $\le$6 arcsec, see Tab. \ref{sample}). These are particularly useful when the radio morphology is complex, or the image angular resolution is not high enough to clearly identify the core position. The following works were considered: \cite{2013ApJS..207...19B,2012MNRAS.426.1750M,2016MNRAS.460...19M,2014A&A...565A...2M,2015MNRAS.451.2370M,2012A&A...545A.101P,2014A&A...561A..67P}.


\section{Morphology}

In this section we discuss the morphology of all the 15 sources in the sample using a combination of the new radio images and images already available in the literature. In particular, for four sources we report for the first time the GMRT maps from our campaign: for J0318+684, IGR J14488-4008, and IGR J17488-2338 at 325 MHz, while at 150 MHz for PKS\,2331$-$240 (see Fig. \ref{GMRT_maps}). In appendix A, the best images available for the remaining sources are presented (Fig. \ref{maps1} and \ref{maps2}), together with the extracted flux densities (both total and for individual components, see Tab. \ref{fluxes}).

\subsection{Comments on individual sources}

\subsubsection*{GMRT images}

\begin{figure*}
  \includegraphics[width=1.0\textwidth]{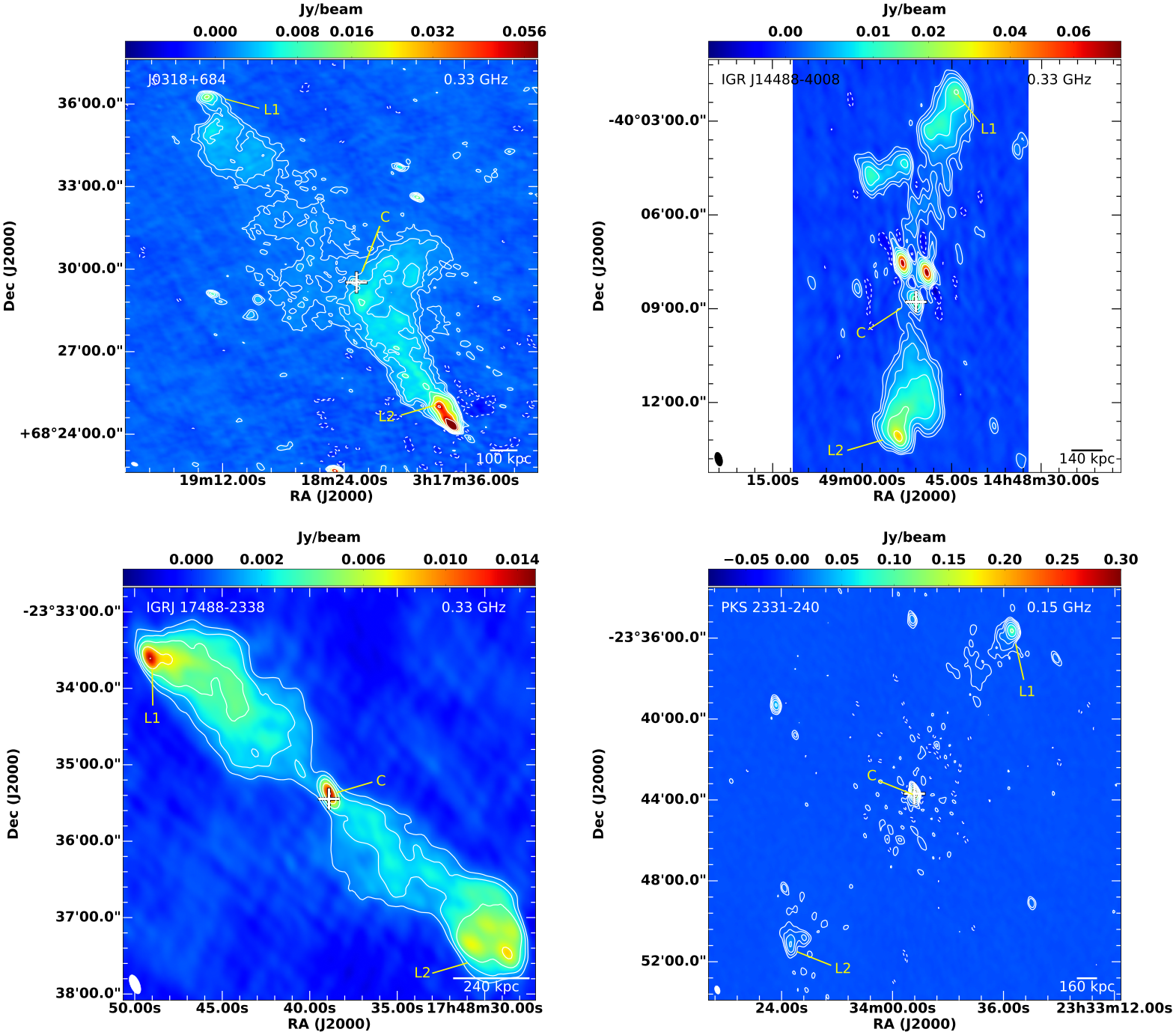}
\caption{GMRT images of the four sources from our campaign. The beam is represented in the bottom-left corner, while the spatial scale in kpc is on the lower-right one. Positions of the cores in X-ray observations from previous works are shown as a white cross. Contours are multiples of the image RMS (3$\sigma$ $\times$ --1, 1, 2, 4, 8, 16, 32, 64) with the first solid line indicating the 3$\sigma$ level.}
\label{GMRT_maps}
\end{figure*}

\begin{itemize}

    \item J0318+684: this source has been reported as GRG by \cite{1998A&A...336..455S}, and is one of the most extended in our sample (1.5 Mpc). Those authors found an inverted-spectrum radio core, proposing this to be suggestive of a Giga-Hertz Peaked Spectrum (GPS) source resulting from a restarted nuclear activity (as later confirmed in \citealt{2019ApJ...875...88B}). Our GMRT map has an angular resolution of 15"$\times$8", at a position angle of 70$\degree$, that is the most detailed image of this source to date. The morphology at 325 MHz shows a diffuse emission starting from the core and extending towards NW, perpendicularly to the axis of the main jets (detected in \citealt{1998A&A...336..455S} as well). This region reaches a distance of 180" from the core, corresponding to 310 kpc at the redshift of this source ($z$=0.09). Thanks to the high sensitivity of our GMRT observations, a fainter counterpart of this emission is visible symmetrically with respect to the core (SE), with a similar extension (see Fig. \ref{0318_zoom} for a zoom on these regions). 
    Thus the source can be ascribed also to the class of X-shaped radio galaxies. This unusual shape can be either the result of a plasma backflow from the lobes, or the remnant of a former activity, where jets were launched at a different angle with respect  to the most extended lobes.  We note  that in the light of dynamical modeling, \cite{2011MNRAS.413.2429M} reported an age 250-270 Myr for the outer lobes of this source. This combined with the finding that the central core has a GPS spectrum, and is therefore young, points to at least two phases of activity, possibly  3 (see Sec. 5.2 for a spectral analysis of this region) if the X-shaped emission is interpreted as an extra phase. However, a more detailed study of the source, including a proper synchrotron spectral ageing analysis, is needed to test this 3-phase hypothesis further.
    
    \item IGR\,J14488--4008 has been presented as a newly discovered GRG in \cite{2015MNRAS.451.2370M}, together with the GMRT map at 610 MHz. Here we present the GMRT map at 325 MHz from the same campaign, confirming the morphology found in \cite{2015MNRAS.451.2370M}. As visible in Fig. \ref{GMRT_maps} (top-right panel), two sources complicate the morphological study of this target: a radio galaxy towards NE, showing a couple of lobes and a core, and another couple of point sources close to the core, $\sim$50 arcsec NW from the map center. The main lobes have a complex structure, presenting hints of an X-shaped structure (as suggested in \citealt{2015MNRAS.451.2370M}), or jet-precession signature, but deeper observations are needed to assess one scenario over the other. Finally, the source shows a GPS core \citep{2019ApJ...875...88B}, which is at odds with the large scale, most probably old, structure. Indeed, taking the formalism of \cite{1999A&A...344....7P}, the lobe magnetic field values reported in \cite{2015MNRAS.451.2370M}, and assuming a break frequency above 1 GHz for the lobes, we can roughly estimate an upper limit to the radiative age of the lobe of around 70 Myr.    
    
    \item IGR\,J17488--2338 is a recently discovered GRG in the NVSS \citep{2014A&A...565A...2M}. In our GMRT map (Fig. \ref{GMRT_maps}), a clear FRII morphology is visible, with bright, predominant lobes and a powerful core connected by fainter emission bridges along the jets. The angular resolution of these GMRT observations (16"$\times$7", at a position angle of 23$^\degree$) has a factor of three improvement with respect to the one from NVSS, allowing us to see in more detail the complex structure of the lobes and better locate the core. From a 2D Gaussian fit of the core region we estimate a deconvolved size of 10"$\times$4", corresponding to a projected linear size of 39$\times$16 kpc, and a integrated flux density of 19$\pm$2 mJy. The lobes structure is complex, and marginally resolved in the NVSS: the Southern lobe presents four different emitting regions, indicative of a complex interaction with the Intergalactic Medium (IGM), while the Northern one shows the typical "bullet"-like shape.   
    A GPS component was found in the core of this source \citep{2019ApJ...875...88B} suggesting a recently restarted activity, not yet visible in the radio morphology in the form of a new couple of jets. Given the above measured extension of the core component, and the expected linear size for GPS sources ($\sim 1$ kpc, \citealt{1998PASP..110..493O}), we can expect that the core region contains a much more complex structure, not resolved in this image, that would deserve further observations at higher frequencies and thus resolution. As a whole, this is an example of restarted radio source that would not be catalogued as such from purely morphological studies at the kpc scale, and whose nature can be unveiled through a spectral study of the core.  
    
    \item PKS\,2331--240: a dedicated study on this source has been presented in \cite{2017A&A...603A.131H,2018MNRAS.478.4634H}. Optically classified as a Seyfert 1.8 from the presence of weak broad emission lines, it shows an unabsorbed X-ray spectrum, typical of AGN seen face-on. Moreover, WISE colours classification corresponds to a BL Lac type \citep{2014ApJS..215...14D}. VLBI observations at different frequencies (8, 15, 24 GHz) revealed a core-jet structure in the nuclear region, indicative of an orientation towards the line of sight, also confirmed by the overall nuclear Spectral Energy Distribution (SED, see \citealt{2017A&A...603A.131H} for a detailed analysis). Further observations in the optical and X-ray band indicated a high flux variability ($\sim$60\% in 7 years), with variable broad H$_{\alpha}$ emission and a broad [OIII] component, suggesting the presence of an outflow \citep{2018MNRAS.478.4634H}. Recently, \cite{2017ApJ...836..174C} presented this source as restarted, finding it to have a convex radio spectrum in the frequency range from MHz to GHz. The GMRT image at 150 MHz presented here confirms the GRG nature of this source, and shows a lack of emission between the lobes and the core region compatible with the change of jet axis proposed by \cite{2017A&A...603A.131H}. 
    
    \end{itemize}

    
\subsubsection*{Images from surveys and the literature}

In the following we discuss the morphology of the other sources, using the best available radio images from surveys or the literature:    
    
    \begin{itemize}

        \item B3\,0309+411B: presented in \cite{1989A&A...226L..13D} as a core-dominated Mpc-sized radio galaxy, this source shows a quasar-like optical spectrum and a strong radio flux density variability for the core ($\sim$35\% in two years), suggesting an orientation with a small angle to the line of sight. WSRT observations at 327 MHz and 608 MHz carried out by those authors show an asymmetric flux intensity between the extended jet and counter-jet, supporting this hypothesis. \cite{2004MNRAS.355..845K} performed GMRT observations at 606 MHz, revealing a curved jet with an extention up to 250 kpc from the core, and then bending towards the northern hotspot. The core spectrum from \cite{2019ApJ...875...88B} is inverted, indicating a possible peak at tens of GHz, and thus suggesting an age even younger than for GPS sources. Compared to the likely old age of the Mpc structure, the core spectrum points to a new radio phase also in this source. In fact, the estimate of the lobe magnetic field as reported by \cite{1999MNRAS.309..100I}  allows a rough estimate of the lobe radiative age as done for the case of IGR J14488-4008: assuming a spectral break above 1 GHz the estimated age is below 10 Myr. Moreover, pc-scale resolution images for this source, collected as part of the 2 cm VLBA survey and Monitoring Of Jets in Active galactic nuclei with VLBA Experiments (MOJAVE, see \citealt{2018ApJS..234...12L} for the latest results), show a one-sided jet extending towards NW, aligned with the jet axis seen on the kpc scale. This confirms the small viewing angle suggested by the optical spectrum and radio variability mentioned above. We report here the TGSS image (Fig. \ref{maps1}, top-left panel), where a core-jet structure compatible with the one in \cite{2004MNRAS.355..845K} is visible, while the counter-jet falls below the noise level.
    
        \item PKS\,0707--35: presents a misaligned double-double morphology (X-shaped) as shown by \cite{2013MNRAS.436..690S}. In Fig. \ref{maps1} (top-right panel), we report the ATCA image at 2.5 GHz presented by those authors: following their interpretation, the almost symmetric emission near the core is the remnant of a previous jet activity, with a different axis with respect to the external and much brighter lobes (sign of a recent fueling of plasma). The jet axis change, that presumably happened at a rate of a few degree/Myr, could likely be the cause of the X-shaped morphology. The same authors provide a radiative age for the inner lobes $<$100 Myr and, based on the high advance speeds of the ends of the outer lobes (0.3$c$), suggest a relatively young age of $\sim$6 Myr for the Mpc structure.
        
        \item 4C\,73.08: thanks to a combination of interferometric and single-dish observations at 1.4 GHz, \cite{2016AcA....66...85W} found an extended radio emission enveloping the whole Mpc-scale structure of the lobes. The map obtained in that work is presented in Fig. \ref{maps1} (middle-left panel). This is in agreement with \cite{1986PhDT.........5J}, who proposed that the hotspots visible in the lobes of this source are more recent, refueled elements, surrounded by an older radio structure. We note that \cite{2011MNRAS.413.2429M} provides a dynamical age for the source lobes between 180 and 230 Myr, implying  a  much older age for the cocoon emission.
            
        \item B2\,1144+35B was previously studied in detail by \cite{1999ApJ...522..101G,2007A&A...474..409G}, making use of multiscale VLA, MERLIN, and VLBA observations (from Mpc to pc-scale). Flux density variability for the arcsec-core up to 50\%, on a 20-years time scale, has been detected by those authors, and interpreted as a possible change in the external medium along the jet. Also, discontinuities between the extended lobes and the arcsec core-jets structure were found and interpreted as the presence of two different phases in the past 10 Myr, involving a possible double-double structure, the inner jets being visible at kpc-scale resolution in their VLA image. \cite{1999ApJ...522..101G} concludes that the extended, Mpc-scale, emission is a relic with an age in the range 50-90 Myr, while the kpc-scale one should be less than 10 Myr old. Finally, they suggest that a merger event could be at the origin of the restarted activity, as suggested by the boxy shape of the optical image. In Fig. \ref{maps1} (middle-right panel), we present the NVSS image where the core is well visible at the center: there are two intervening sources - a point-like one north of the core, and a radio galaxy NW of the upper-right lobe - complicating the morphology. 
          
        \item HE\,1434--1600: this source is the most extended GRG of our sample, with a projected linear size of 1.8 Mpc. The NVSS image (Fig. \ref{maps1}, bottom-left panel) clearly shows an asymmetric, possibly core-jet, structure between the two extended lobes: a single-jetted morphology is typically present in radio sources with a jet axis close to the observer's line of sight, causing a relativistic Doppler-boosting that enhances the emission of the jet coming towards the observer, and dims the one of the counter-jet. This could be the case for the core-jet structure of this source, implying an inner jet axis within $\sim$45$^\circ$ from the line of sight, while the symmetric morphology of the relaxed lobes does not suffer from this relativistic effect. The inner jet orientation at a small angle from the line of sight is also supported by the optical classification of the core (Broad Line QSO, see \citealt{2016MNRAS.461.3165B}). Finally, \cite{2004A&A...424..455L}, through VLT optical spectroscopy, found that the host of this AGN is part of a group of at least five galaxies, and most probably underwent a collision with its nearest companion ($\sim$4 arcsec). Later, \cite{2015ApJ...804...34H} confirmed the merging state thanks to deep optical imaging. As a whole, this source could have undergone a restarted event triggered by the merger, but further observations are needed to confirm this scenario. 
            
        \item 4C\,+63.22: the TGSS image for this source (Fig. \ref{maps1}, bottom-right panel) shows an external couple of lobes, and an inner diffuse emission with a different axis with respect to the kpc-scale structure, similarly to X-shaped sources. This could be tentatively interpreted as a inner couple of ancient lobes, pertaining to a previous radio phase, but also as a distortion of the relaxed plasma due to its buoyancy in the intergalactic medium. A clear conclusion cannot be drawn at this angular resolution and sensitivity. The detected morphology is more evident in the higher resolution Low-Frequency Array (LOFAR) image, subject of a future work (Bruni et al. in prep.). If the X-shaped morphology is confirmed, this source could be a case similar to PKS\,0707--35 and J0318+684, where the remnants of the previous radio activity are visible between the more recent and extended lobes.
    
        \item Mrk\,1498 shows a couple of strong lobes, and a well defined core, but the jets connecting them are barely visible in the TGSS images (Fig. \ref{maps2}, top-left panel). The core shows a typical GPS radio spectrum, indicating a recently restarted radio activity \citep{2019ApJ...875...88B} that suggests the presence of two different phases.   
        A multi-wavelength study of this source, showing signs of a possible reactivation as the consequence of a merger event, was presented in \cite{2019MNRAS.489.4049H}. In that work, a pc-scale VLBA image of the radio core region, showing recently emitted jet components, is presented, thus confirming the ongoing activity linked to the central GPS radio source.  
    
        \item 4C\,+34.47 presents bright lobes dominating the emission in the TGSS maps (FRII morphology, Fig. \ref{maps2}, top-right panel). \cite{2010A&A...523A...9H}, through detailed VLA observations, detected a one-sided jet in the inner kpc, analogously to 4C\,+63.22. An estimate of the viewing angle from the jet/counterjet emission ratio and kinematic arguments led to values $<$30$\degree$. This would imply a deprojected linear size in the range 1.0--3.4 Mpc, placing this source among the most extended radio galaxies. The optical classification of this source's core supports a line of sight close to the jet axis (Seyfert 1). Also for this source, an estimate of the lobe magnetic field as reported by \cite{1999MNRAS.309..100I} allows a rough estimate of the Mpc lobe radiative age as done for the case of IGR J14488-4008 and B3 0309+411B: in this case we also assume a spectral break above 1 GHz as suggested by the steep spectral index measured by \cite{2010A&A...523A...9H}. The estimated age is below 11 Myr, again at odds with the finding of \cite{2019ApJ...875...88B} who reported a GPS core in this source, and suggested a recent reactivation of the jet.
       
        \item PKS\,2014--55: this source is atypical in many ways. It is a clear example of an X-shaped radio galaxy,  the first to be found in an FR I type source, but also the only one associated to a GRG. We report here the ATCA map at 1.4 GHz from \cite{2008arXiv0806.3518S}, showing  a double-double radio structure with an inner couple of lobes (see Fig. \ref{maps2}, middle-left panel and relative zoom), aligned with the 1.5 Mpc outer lobes and nearly two orders of magnitude smaller in linear extent. This second more recent phase of activity  is thus embedded in the Mpc-scale (and thus older) emission region. Those authors suggested that the X-shaped morphology of the more extended structure could be due to interaction of a gas backflow with the host galaxy halo.
        
        \item 4C\,+74.26: the TGSS image reported here (Fig. \ref{maps2}, bottom-left panel) shows a typical FRII morphology, with prominent emission from the lobes. It is one of the largest radio sources associated with a quasar. \cite{1992MNRAS.259P..13P} revealed a one-sided, pc-scale, nuclear emission through VLBI observations, well aligned with the kpc-scale jet visible with the VLA: this is suggestive of a jet orientation close to the line of sight, with an estimated angle $<$49$^\circ$. Finally, the core is reported to show a GPS radio spectrum in \cite{2019ApJ...875...88B}, while the lobes have an age of 90-120 Myr as estimated from dynamical considerations \citep{2011MNRAS.413.2429M}.
        
        \item PKS\,2356--61: this source was observed with ATCA at 1.4 GHz by \cite{1996MNRAS.279..257S}, who detected two unresolved hotspots at the lobes edges, connected by continuous emission peaking into two diffuse emission regions symmetrically located about the core.  Moreover, those authors noted a minimum of emission along the jets axes, reminding a double-double morphology. A similar morphology is seen in the ATCA image at 20 GHz presented by \cite{2009MNRAS.395..504B}, where a further minimum of emission, this time between the core and the supposed inner sets of lobes, is evident. Moreover, the bottle-necked morphology of the external sets of lobes (Fig. 8 in \citealt{1996MNRAS.279..257S}), could be a consequence of the jets emerging from a past activity emission region (see discussion in \citealt{2013MNRAS.436..690S}). The latter could be traced by the electric field vectors map as presented by \cite{2009MNRAS.395..504B} (see their Fig. 8), showing a tentative circumferential magnetic field structure, which is also seen for the weaker inner pair of lobes suggestive of possible termination points associated with past activity. 
        In the SUMSS image shown in Fig. \ref{maps2} (bottom-right panel, at lower resolution with respect to the mentioned ATCA observations) two sets of lobes are visible. Finally, a High Frequency Peaker (HFP) core has been found by \cite{2019ApJ...875...88B}. As a whole, this could be an example of a GRG with a frequently recurring radio activity, that most probably went through more than two radio phases in its history.
    
\end{itemize}


\section{Signs of restarted radio activity}
\label{radio}

Nine sources of our sample presents signs of restarted radio activity from previous works in the literature (see Tab. \ref{sample} for details). Another four are found to show hints of a restarted radio activity from the Effelsberg-100m campaign presented in \cite{2019ApJ...875...88B}. In the following, we divide the sources into the main groups presenting evidence of a restarted nuclear activity.


\subsection{GPS cores}
In \cite{2019ApJ...875...88B}, we have shown that 61\% of sources from this sample have a GPS/HFP-like spectrum for their core region (see Tab. \ref{sample} for details). With the latter being commonly considered as an indicator of a young radio phase \citep{1998PASP..110..493O,2003PASA...20...79D}, this suggested a restarted radio activity, since GPS sources have usually an age of a few kyr, while the Mpc-scale structure of these GRG took tens of Myr to reach such an extent. Indeed, using simple kinematic arguments and assuming  a typical velocity of advancement of 0.1$c$, the outer lobes would traverse a distance of 0.7-1.8 Mpc (the range of sizes observed in our sample) in  20-60 Myr . These values are fully compatible with age estimates reported here for individual sources either collected from the literature or based on well known arguments. Albeit these estimates are affected by large uncertainties and are based on various  assumptions, nevertheless provide an indication that the Mpc structure in our GRG sample is indeed Myr old. 

The maps collected for this work allow us to study the morphology of these GRG hosting a young radio nucleus. Five among them are hosted by typical FRII GRG, with predominant emission from the lobes (IGR J14488-4008, Mrk 1498, 4C\,+34.47, IGR\,J17488-2338, 4C\,+74.26): this is a fraction of objects that would not be recognized as restarted from their radio morphology only, thus hinting to a hidden fraction of restarted RG in radio imaging catalogues such as FIRST, NVSS, SUMSS. Another two show possible signatures of an X-shaped or double-double morphology (J0318+684 and PKS 2356-61, respectively), in which the restarted core could indicate a third radio activity phase. We discuss the possible implications for the radio duty cycle in Sec. \ref{discussion}. Finally, one source (B3 0309+411B) shows no clear indication of a restarted jet on the Mpc scale. A more detailed map for this source obtained with LOFAR data will be presented in a future paper (Bruni et al. in prep.), in which the counterlobe is clearly detected, thus confirming the two lobes morphology for this source. 

Previous authors raised concerns about the reliability of a GPS classification on the only basis of a convex spectrum built with sparse data \citep{2005A&A...435..839T,2007A&A...469..451T}. In particular, they pointed out how - specially in the case of a Blazar-type GPS - a flaring state could mimic a GPS-like spectrum through a newly ejected component. They proposed, as a viable solution to this, to collect simultaneous multi-frequency data to define the turnover, and to perform long-term monitoring to test any variability-induced spectral shape.
For the classification presented in \cite{2019ApJ...875...88B}, we collected simultaneous flux density measurements with the Effelsberg 100-m telescope, thus providing reliable data between 4 and 10 GHz. Moreover, all of ours GPS cores are optically classified as Seyferts in the literature (see \citealt{2016MNRAS.461.3165B} for a summary), possibly excluding a flare-induced convex spectrum shape. Nevertheless, three of them showed radio variability in previous studies (4C +74.26, 4C +34.47, and B3 0309+411) that could play a role in their spectral shape. Our ongoing high-resolution campaign on this sample will allow us to verify the jet orientation of these cores, possibly disentangling their nature.


\begin{figure*}
    \includegraphics[width=160mm]{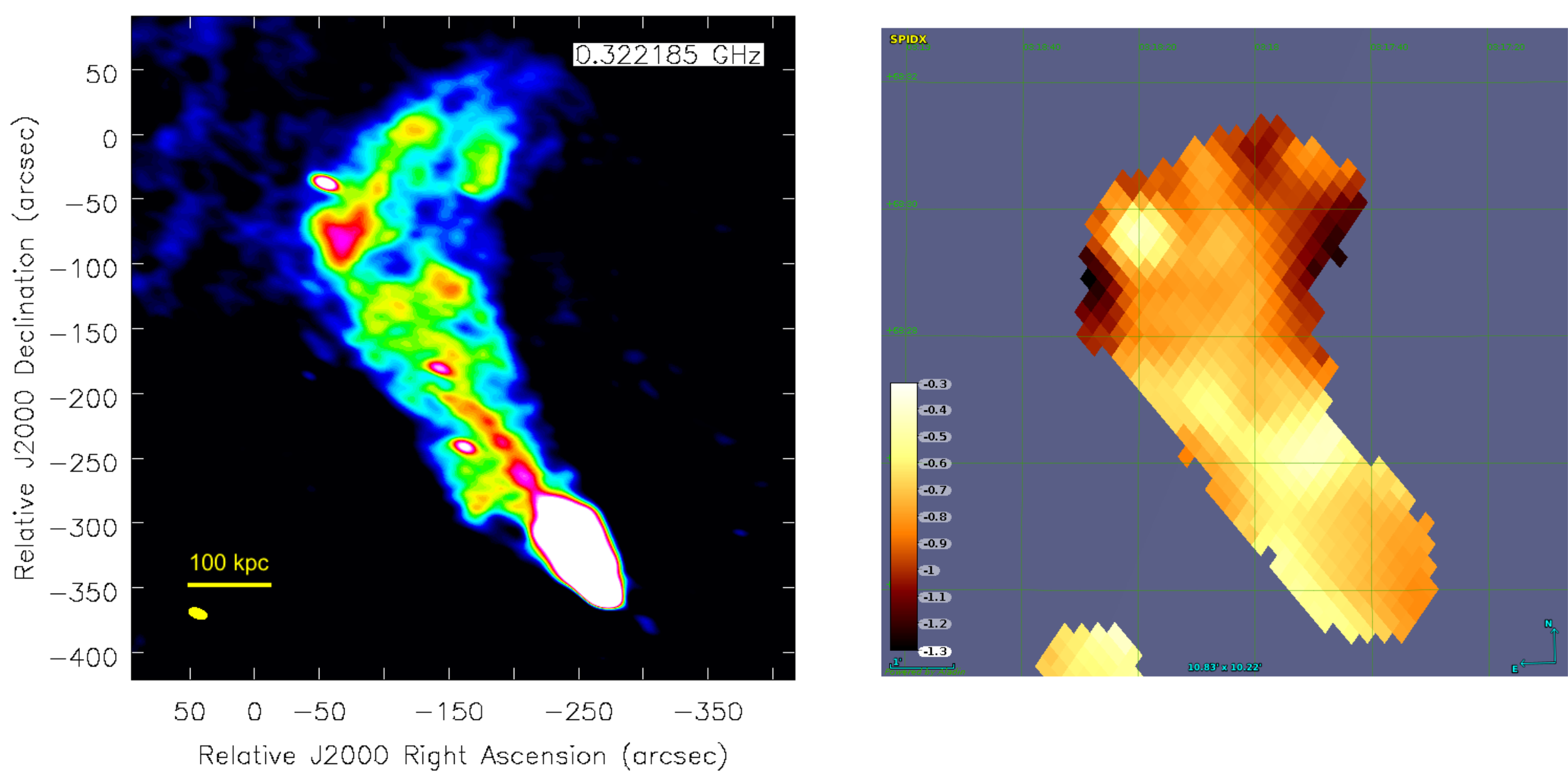}
    \caption{Left panel: zoom on the Southern region of source J0318+684, the image is obtained from GMRT data at 325 MHz. The beam is reported in yellow in the lower-left corner. Right panel: spectral index maps between 150 MHz and 1.4 GHz for the same region, from the SPIDX catalogue \citep{2018MNRAS.474.5008D}.}
    \label{0318_zoom}
\end{figure*}


\subsection{Double-double and X-shaped GRG}

Three among our GRG show two sets of lobes, suggesting the presence of two distinct radio activity phases (PKS\,0707--35, PKS\,2014--55, and possibly PKS\,2356--61). These were also the subject of previous detailed studies, realized with ATCA data in the GHz domain (\citealt{2008arXiv0806.3518S,2013MNRAS.436..690S,1996MNRAS.279..257S}). From their work, the emission corresponding to the off-axis inner lobes could also be due to other mechanism, such as plasma back-flowing from the lobes towards the nucleus.

Moreover, another three sources have a morphology that is suggestive of the same characteristic: J0318+684, B2\,1144+35B, and 4C\,+63.22. The first one shows a symmetric, off-axis emission directly linked to the core region. \cite{2018MNRAS.474.5008D} presented a spectral index catalogue built from TGSS and NVSS images (thus between 150 MHz and 1.4 GHz), covering $\sim$80\% of the sky, at a declination $>-40^{\circ}$. For J0318+684, the angular resolution of the spectral index map provided by the catalogue is good enough to allow a characterization of the emission coming from the off-axis regions. As visible in Fig. \ref{0318_zoom} (right panel), a very steep spectral index ($<-1$, with an error $<$20\%), usually corresponding to relic emission regions at much larger distance from the core, is estimated for the area corresponding to the candidate inner couple of lobes. The values are in agreement with the ones previously obtained by \cite{1998A&A...336..455S} using Westerbork Northern Sky Survey at 327 MHz (WENSS, \citealt{1997A&AS..124..259R}) and NVSS data, but at lower angular resolution. This might suggest that these regions pertain to a previous activity phase of the source, and what we are seeing is the remnant of the lobes formed in that epoch. The misalignment with respect to the most recent jet axis could hide the real extension of this structure, that eventually could have a de-projected linear size comparable with the most recent radio structure in case of small angle to the line of sight. 

The second mentioned source, B2\,1144+35B, has a more complex morphology. From the TGSS map presented in Fig. \ref{maps2}, it is not possible to study in detail the structure of the jet, so we rely on the previous multi-scale studies by \cite{1999ApJ...522..101G,2007A&A...474..409G}. From their work, discontinuities between lobes and inner kpc-structure were found, possibly hinting to the presence of two distinct radio phases. However, dedicated observations with the upgraded JVLA could probably provide better evidence of such characteristics. Thus, we tentatively consider this GRG as a double-double radio source. 

Finally, we discuss here for the first time the TGSS image of 4C\,+63.22, that shows a partially resolved X-shaped structure. Indeed, an off-axis emission between the more extended lobes, apparently linked to the core, is visible in Fig. \ref{maps1}. No previous records in the literature were found for this source. In a future paper, presenting follow-up LOFAR data, we will better define and clarify the inner structure of this GRG (Bruni et al. in prep.). 

A further GRG of our sample is most probably undergoing a new radio phase, but with the new jets axis misaligned by about 90$^\circ$ with respect to the previous ones: PKS\,2331--240. Indeed, \cite{2017A&A...603A.131H} showed that the core of this GRG presents typical properties of a BL Lac, both in terms of radio morphology at pc-scale resolution, and overall multi-wavelength spectral properties. Thus, this could be a case of X-shaped GRG, in which the new jet axis lies at a small angle to the line of sight, thus not showing the typical "X" morphology.

Among leading scenarios for X-shaped RG is one where the jets get activated in a new direction after suffering a large-angle spin flip in a binary SMBH system (see e.g. \citealt{2018ApJ...852...48S} and references therein), produced by a major galaxies merger. Nevertheless, other authors propose that this peculiar morphology could be the result of hydrodynamic processes. Indeed, \cite{2019MNRAS.488.3416H} recently presented Low-Frequency Array (LOFAR) observations of NGC\,326, a prototypical X-shaped radio galaxy, showing that high-sensitivity imaging in the MHz domain highlight a large-scale structure enclosing the previously detected X-shaped emission. Requiring a merger-related hydrodynamic process, the presence and morphology of these extended wings underline how the presence of a binary BH is not the only factor determining the X-shape in some radio galaxies. Previously, \cite{2016A&A...587A..25G} performed a combined optical/radio study of 53 X-shaped RG, finding that the off-axis radio emission is often aligned with the host galaxy minor axis. This provides support to a scenario in which plasma expands in a asymmetric external medium, that eventually produce the observed morphology by hydrodynamic processes. In the light of these previous works, we cannot exclude that the X-shaped morphology found in our sample is the result of such processes, without directly implying a restarted radio phase. Nevertheless, our previous X-ray follow-up on this sample \citep{2018MNRAS.481.4250U} highlighted how the radio lobes luminosities and jet powers are low when compared with the core X-ray emission: this could partially support a restarted scenario in our X-shaped objects.

In the literature, the fraction of X-shaped objects among the radio galaxies population is only of a few percent (4-7\%, \citealt{2018ApJ...852...47R,2018ApJ...852...48S}): in our sample of hard X-ray selected GRG, despite the reduced statistics, we find it to be more than twice that fraction ($\sim$13-25\% considering the 4 objects discussed above). On the one hand, the fact that we are considering only objects with noticeable radio linear size obviously can play a role, allowing more time for a possible merger to happen during the considerably long activity period of our objects. On the other hand, the first selection step of our sample, i.e. a high hard X-ray luminosity (typically larger than $10^{43}$ erg/s) has the effect of selecting highly accreting objects: this could favour the presence of objects that recently underwent a merger (providing abundant accreting material) and thus more easily harbouring a binary BH system with respect to lower luminosity AGN. 
This hypothesis will be tested with the recently obtained European VLBI Network (EVN) follow-up observations of the four X-shaped GRG candidates (Bruni et al. in prep.). 


\subsection{Radio Cocoon}

A further method to reveal previous phases of radio activity is the detection of diffuse, relic emission in the surroundings of the radio galaxy. Given the low surface brightness of this "cocoon" emission, it can easily be missed at the angular resolution of interferometric observations, dropping below the image noise. Given the large angular size, also the sampling of short baselines becomes critical. It is possible to overcome this issue by filling the zero-baseline visibilities gap of the interferometer with ad hoc single-dish observations, that largely improves the reconstruction of such structure in the image plane. The final image is then able to provide both high angular resolution details, and sensitivity to extended structures. This observing technique was adopted by \cite{2016AcA....66...85W} to study a sample of 15 GRG at 1.4 GHz, combining Effelsberg 100-m single-dish observations with NVSS data. They found that 5/15 sources showed the presence of an extended radio cocoon enveloping the Mpc-scale structure. The overlap with our sample is of 3 sources (J0318+684, 4C\,73.08, and Mrk\,1498), among which only one (4C\,73.08) shows the presence of such extended emission.
We do not exclude that, by applying this technique to the whole sample, we could identify more sources like this one. In particular, LOFAR could provide at the same time high-sensitivity and high-resolution images of these GRG, making it possible to identify diffuse, low surface brightness emission.


\section{Conclusions}
\label{discussion}

This work, presenting the morphology of the GRG sample from \cite{2016MNRAS.461.3165B}, is the third in a series after \cite{2018MNRAS.481.4250U} and \cite{2019ApJ...875...88B} aiming at characterizing the combined X-ray and radio properties of the first GRG sample selected from \emph{INTEGRAL}/IBIS and \emph{Swift}/BAT hard X-ray catalogues ($>$20 keV).
\cite{2016MNRAS.461.3165B}, who presented the parent sample of these GRG, found that this way of selecting radio galaxies favours large-sized objects, since $\sim$60\% of their sample of radio galaxies emitting at hard X-rays has a size larger than 0.4 Mpc, and the fraction of GRG (22\%) is about four times larger than what seen in radio surveys to date ($\sim$6\%, \citealt{1999MNRAS.309..100I}). From our previous follow-up study in the X-ray domain, the nuclei of these GRG already showed an excess of nuclear X-ray emission with respect to the lobes radio luminosities and estimated jets power, indicating a possible restarted nuclear activity \citep{2018MNRAS.481.4250U}. Furthermore, a first single-dish radio follow-up conducted with the Effelsberg 100-m telescope, and aiming at building the radio spectra of the cores of these GRG, revealed the presence of young radio sources in 8 over 15 targets. The typical age of such sources being of the order of a few kyr, in contrast with the much more extended Myr time-scale structure. This suggested a restarted event in these GRG \citep{2019ApJ...875...88B}.

In this paper, we presented an analysis of the radio morphology from the low frequency images of the 15 GRG, collected either from surveys, the literature, or our GMRT campaign. We conclude that at least 13 over 15 of these sources show features consistent with a restarted radio activity. Indeed, eight sources host a GPS/HFP core from our previous study \citep{2019ApJ...875...88B}, six show a double-double or X-shaped morphology (two among these have also a GPS core), one underwent a jet reorientation of about 90 degrees, with the new jet pointing a few degrees from the line of sight, and finally another one is enveloped in a diffuse radio emitting cocoon, relic of a previous activity phase. 
Moreover, two (plus another two with more marginal evidence) over 15 sources present a X-shaped morphology, and are possible candidates for hosting a binary supermassive BH system. To summarise, at least 13 over 15 of the sources presented here show features which are consistent with the possibility of restarted radio activity. As a whole, our results highlight the peculiarity of this GRG sample selected from hard X-ray catalogues, indicating that such selection could favour the discovery of particularly active jetted AGN, able to form Mpc-scale structures and undergo multiple radio activity episodes. Further studies to confirm the restarted activity in these GRG through high sensitivity imaging and spectral ageing analysis are in progress and will be presented in future works.

\subsubsection*{Final remarks}

To date, a systematic study of the radio duty cycle on large samples of RG is still missing, given the sensitivity and resolution needed to study in details these complex structures. In the literature, a few works on duty cycle are present. Among recent ones, \cite{2013MNRAS.430.2137K} studied two double-double RG (J0116--4722 and J1158+2621), concluding that the eldest couple of lobes are the result of an active phase that lasted $\sim$100 Myr on average, with a quiescent period of about 1 Myr before the AGN reignited into the second phase. \cite{2018A&A...618A..45B} investigated the spectral properties of the outer lobes of another known double-double RG (B2 0258+35), finding that either they are still fueled, or the jets feeding them switched off less than $\sim$10 Myr ago. In other cases, a longer pause between the two phases has been found ($\sim$100 Myr, \citealt{2007MNRAS.378..581J}). Our GRG sample, given the large fraction of restarted objects and the resolved morphology, offers the opportunity to have at hand AGN that are undergoing different stages of the restarted activity, and can thus be an interesting laboratory to test the radio duty cycle in AGN. 

\section*{Acknowledgements}
We thank M. We{\.z}gowiec, M. Jamrozy, and K.-H. Mack for providing us with the combined Effelsberg/NVSS map for one of our sources.
G.B. acknowledges financial support under the INTEGRAL ASI-INAF agreement 2019-35-HH.0.
M.B. acknowledges support from the ERC-Stg DRANOEL, no 714245, and from INAF under PRIN SKA/CTA FORECaST.
L.H.G. acknowledges support from FONDECYT through grant 3170527, and partly from Conicyt PIA ACT 172033.
E.C. acknowledges the National Institute for Astrophysics (INAF) and the University of Rome - Tor Vergata for the PhD scholarship in the XXXIII PhD cycle.
This publication has received funding from the European Union's Horizon 2020 research and innovation programme under grant agreement No 730562 [RadioNet].



\bibliographystyle{mnras}
\bibliography{GRG_references} 



\appendix

\section{Collected images and quantities}

In the following we present the images for the 11 GRG having data available from surveys (TGSS, NVSS, SUMSS), or from previous works in the literature. For each source, we present quantities measured from the maps in Tab. \ref{fluxes}.

\onecolumn
\begin{longtable}{cccccccccccc}
\caption{Quantities for the GRG sample presented in this work. For each source we report: projected linear size (LS), telescope and frequency of the image considered here, its RMS, beam, measured flux density for each component, and total radio power. The adopted nomenclature for the different components is: C for the core, L1 for the uppermost lobe, L2 for the lowermost lobe, L3 and L4 for the inner couple of lobes (when present), following the same criterion. The core flux density is not reported in case it could not be identified from the available map.}\\
  \hline
    Source      		 & LS    & Telescope &   Frequency & RMS        &  Beam           &  Component    &   Flux density &  Total Radio Power \\
                		 & (Mpc) &           &   (MHz)     & (mJy/beam) &  (arcsec)       &               &   (mJy)        &  (W/Hz)            \\
    \hline 	
    \endhead
    \hline
    \endfoot
	B3 0309+411B	    & 1.15  &  TGSS     &	150		  & 6.3	  & 25$\times$25    &	 Tot		  &	   748$\pm$75  	&	25.50  		  \\   
						&		&           &			  &		  &                 &	 C            &	   446$\pm$45  	&				  \\
						&		&           &			  &		  &                 &	 L1           &	   128$\pm$13  	&				  \\
	\hline           
	J0318+684           & 1.52  &  GMRT     &	325       & 0.3	  &  15$\times$8    &	 Tot	      &	  3260$\pm$30  	&	25.81   	  \\
						&		&           &			  &		  &                 &	 C            &	  	12$\pm$1   	&				  \\
						&		&           &			  &		  &                 &	 L1           &	   422$\pm$42  	&				  \\
						&		&           &			  &		  &                 &	 L2           &	   976$\pm$98  	&				  \\
	\hline           
	PKS 0707--35 	    & 1.02  &  ATCA    &	2500      & 0.2	  &  18$\times$14    &	 Tot		  &	  1130$\pm$110 	&	25.53   	  \\
						&		&          &			  &		  &                  &	 C            &	  	45$\pm$41   	&				  \\
						&		&          &			  &		  &                  &	 L1           &	   253$\pm$25  	&				  \\
						&		&          &			  &		  &                  &	 L2           &	   565$\pm$56  	&				  \\	
						&		&          &			  &		  &                  &	 L3           &	   132$\pm$13  	&				  \\
						&		&          &			  &		  &                  &	 L4           &	   122$\pm$12  	&				  \\
	\hline           
	4C 73.08		    & 0.94  &  NVSS     &	1400      & 1.0	  &  45$\times$45    &	 Tot		  &	  4340$\pm$430 	&	25.58         \\
						&		&           &			  &		  &                  &	 C            &	    14$\pm$2	&				  \\
						&		&           &			  &		  &                  &	 L1           &	  1100$\pm$110 	&				  \\
						&		&           &			  &		  &                  &	 L2           &	  2080$\pm$210 	&				  \\
	\hline           
	B2 1144+35B		    & 1.20 	&  NVSS       &	1400      & 0.5	  &  45$\times$45    &	 Tot		  &	   784$\pm$78	&	24.84   	  \\
						&		&             &			  &		  &                  &	 C            &	   658$\pm$66 	&				  \\
						&		&             &			  &		  &                  &	 L1           &	    21$\pm$2	&				  \\
						&		&             &			  &		  &                  &	 L2           &	    97$\pm$9	&				  \\	
	\hline           
	HE 1434--1600   	& 1.82	& NVSS       &	1400      & 0.5	  &  45$\times$45    &	 Tot		  &	   229$\pm$23  	&	25.05   	  \\
						&		&            &			  &		  &                  &	 C            &	  	75$\pm$7	&				  \\
						&		&            &			  &		  &                  &	 L1           &	   	34$\pm$3   	&				  \\
						&		&            &			  &		  &                  &	 L2           &	   	94$\pm$9   	&				  \\	
						&		&            &			  &		  &                  &	 L3           &	   	19$\pm$2  	&				  \\
	\hline           
	IGR J14488--4008    & 1.54	& GMRT        &	325       & 0.3	  &  27$\times$14    &	 Tot		  &	   559$\pm$56  	&	25.30  		  \\
						&		&             &			  &		  &                  &	 C            &	    19$\pm$2   	&				  \\
						&		&             &			  &		  &                  &	 L1           &	   158$\pm$16	&				  \\
						&		&             &			  &		  &                  &	 L2           &	   338$\pm$34  	&				  \\
	\hline           
	4C +63.22   	    & 0.71	& TGSS        &	150       & 3.8	  &  25$\times$25    &	 Tot		  &	  3600$\pm$360	&	26.57         \\
						&		&             &			  &		  &                  &	 C            &	  	 -  	  	&				  \\
						&		&             &			  &		  &                  &	 L1           &	  2280$\pm$230 	&				  \\
						&		&             &			  &		  &                  &	 L2           &	   597$\pm$60   &				  \\	
						&		&             &			  &		  &                  &	 L3           &	   287$\pm$29 	&				  \\
						&		&             &			  &		  &                  &	 L4           &	   367$\pm$37   &				  \\
	\hline           
	Mrk 1498  		    & 1.20	& TGSS        &	150       & 2.9	  &  25$\times$25    &	 Tot		  &	  3240$\pm$320	&	25.29   	  \\
						&		&             &			  &		  &                  &	 C            &	  	47$\pm$5 	&				  \\
						&		&             &			  &		  &                  &	 L1           &	   814$\pm$81  	&				  \\
						&		&             &			  &		  &                  &	 L2           &	  2050$\pm$200  &				  \\
	\hline           
	4C +34.47      	    & 0.82	& TGSS       &	150       & 2.6	  &  25$\times$25    &	 Tot		  &	  5800$\pm$580	&	26.82  		  \\
						&		&            &			  &		  &                  &	 C            &	   203$\pm$20	&				  \\
					    &		&            &			  &		  &                  &	 L1           &	  2430$\pm$240 	&				  \\
						&		&            &			  &		  &                  &	 L2           &	  2690$\pm$270 	&				  \\
	\hline           
	IGR J17488--2338	& 1.41	& GMRT        &	325       & 0.3	  &  16$\times$7     &	 Tot		  &	   449$\pm$45  	&	25.83   	  \\
						&		&             &			  &		  &                  &	 C            &	   	15$\pm$2	&				  \\
						&		&             &			  &		  &                  &	 L1           &	   162$\pm$16   &				  \\
						&		&             &			  &		  &                  &	 L2           &	   125$\pm$12   &				  \\
	\newpage
	PKS 2014--55 	    & 1.48	& ATCA        &	1400      & 0.1	  &  7$\times$5      &	 Tot		  &	  2290$\pm$230	&	25.3   		  \\
						&		&             &			  &		  &                  &	 C            &	   	-      	    &				  \\
						&		&             &			  &		  &                  &	 L1           &	   483$\pm$48   &				  \\
						&		&             &			  &		  &                  &	 L2           &	   380$\pm$38   &				  \\
						&		&             &			  &		  &                  &	 L3           &	    49$\pm$5    &				  \\
						&		&             &			  &		  &                  &	 L4           &	   179$\pm$18   &				  \\
	\hline           
	4C +74.26   	    & 1.21	& TGSS        &	150       & 4.1	  &  25$\times$25    &	 Tot		  &	  6780$\pm$680  &	26.23  		  \\
						&		&             &			  &		  &                  &	 C            &	   129$\pm$13	&				  \\
						&		&             &			  &		  &                  &	 L1           &	   962$\pm$96   &				  \\
						&		&             &			  &		  &                  &	 L2           &	  4460$\pm$450  &				  \\
	\hline           
	PKS 2331--240       & 1.17	& GMRT        &	150       & 1.1	  &  25$\times$15    &	 Tot		  &	  1240$\pm$120  &	24.88   	  \\
						&		&             &			  &		  &                  &	 C            &	   314$\pm$31	&				  \\
						&		&             &			  &		  &                  &	 L1           &	   462$\pm$46   &				  \\
						&		&             &			  &		  &                  &	 L2           &	   294$\pm$29   &				  \\
	\hline           
	PKS 2356--61  	    & 0.72	& SUMSS        &    843       & 37	  &  43$\times$49    &	 Tot		  &	 42000$\pm$420	& 26.92	 	    \\
						&		&              &			  &		  &                  &	 C            &	   	-	        &			    \\
						&		&              &			  &		  &                  &	 L1           &	 12900$\pm$1300	&			    \\
						&		&              &			  &		  &                  &	 L2           &	  5410$\pm$540  &		        \\
						&		&              &			  &		  &                  &	 L3           &	  7350$\pm$730  &				\\
						&		&              &			  &		  &                  &	 L4           &	 12900$\pm$1300 &				
\label{fluxes}
\end{longtable}
\twocolumn


\begin{figure*}
 \includegraphics[width=0.9\textwidth]{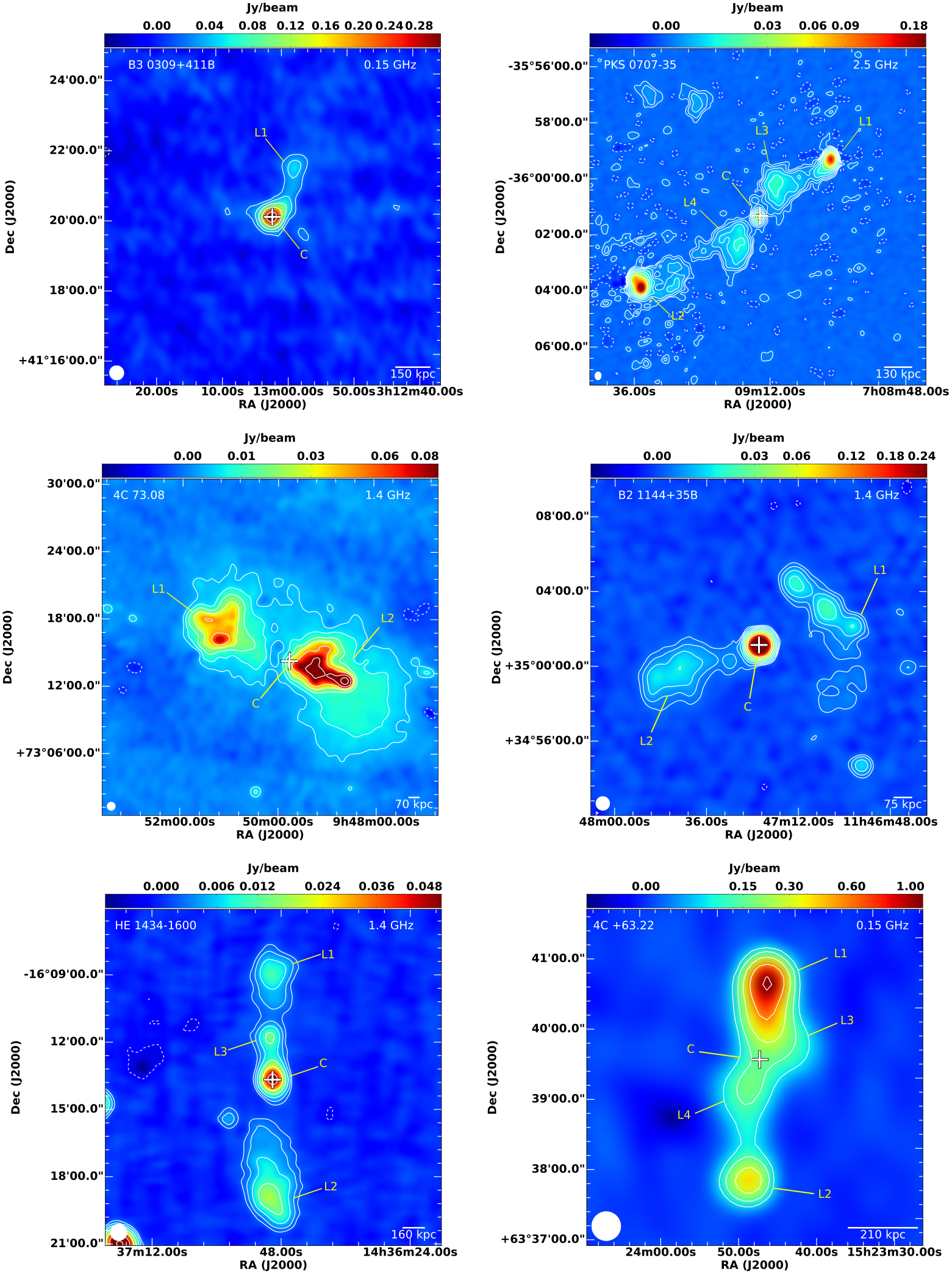}
\caption{Best images available from the literature or previous works for our sources. The beam is reported as a white ellipse on the lower-left corner, while the spatial scale in kpc is on the lower-right one. Positions of the cores in X-ray observations from previous works are shown as a white cross. Contours are multiples of the image RMS (3$\sigma$ $\times$ --1, 1, 2, 4, 8, 16, 32, 64) with the first solid line indicating the 3$\sigma$ level.}
\label{maps1}
\end{figure*}


\begin{figure*}
 \includegraphics[width=0.92\textwidth]{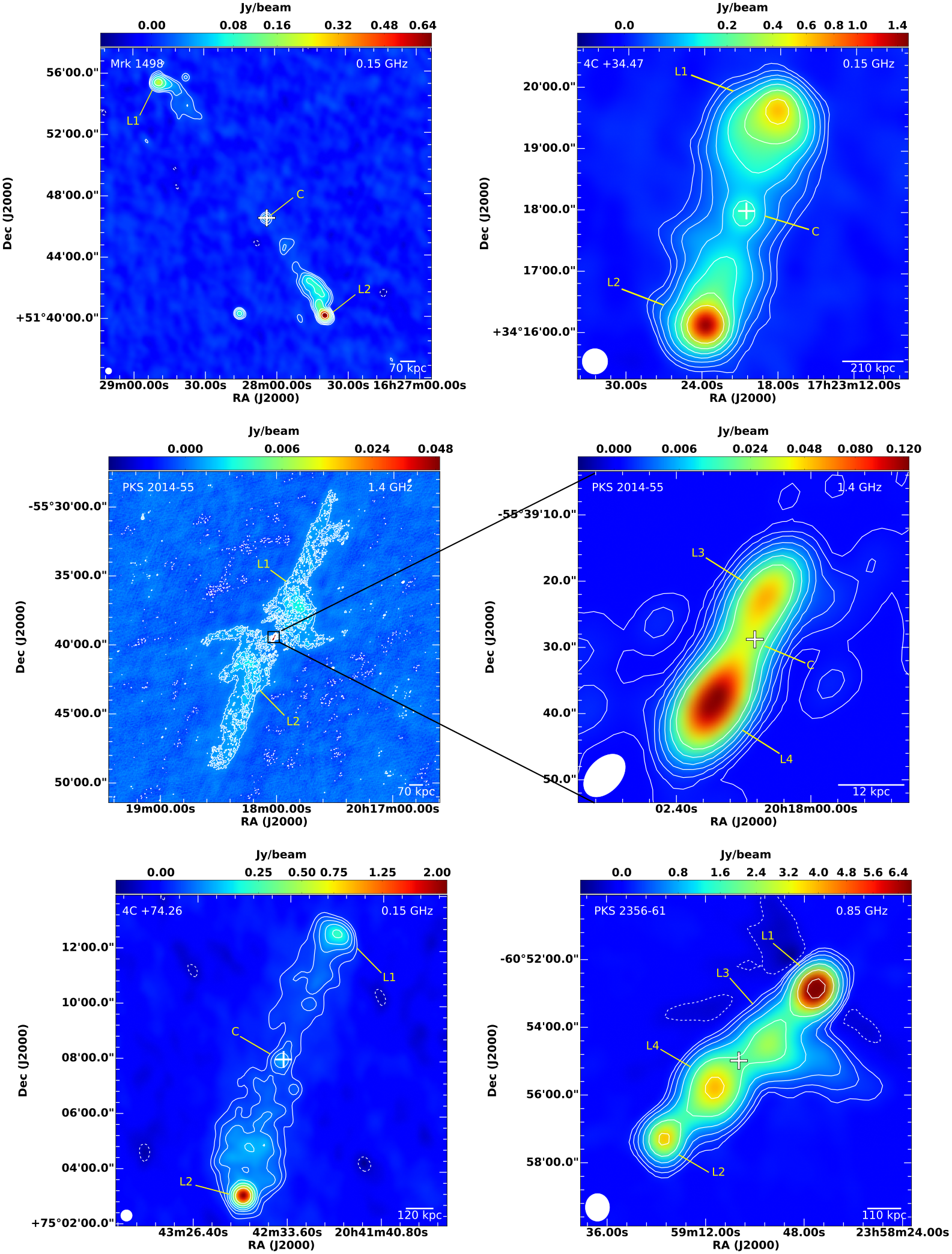}
\caption{Same as for figure \ref{maps1}}
\label{maps2}
\end{figure*}


\bsp	
\label{lastpage}

\end{document}